\documentclass[
 reprint,
 superscriptaddress,
 amsmath,amssymb,
 aps,
 prx,
]{revtex4-1}

\usepackage{graphicx}
\usepackage{dcolumn}
\usepackage{bm}
\usepackage{braket}
\usepackage{color}
\usepackage{physics}

\makeatletter
\def\convertto#1#2{\strip@pt\dimexpr #2*65536/\number\dimexpr 1#1}
\makeatother

\begin{document}

\title{Pulsed Rabi oscillations in quantum two-level systems: beyond the Area Theorem}

\author{Kevin A. Fischer}
\email{kevinf@stanford.edu}
\affiliation{E. L. Ginzton Laboratory, Stanford University, Stanford CA 94305, USA}
\author{Lukas Hanschke}
\affiliation{Walter Schottky Institut, Technische Universit\"at M\"unchen, 85748 Garching bei M\"unchen, Germany}
\author{Malte Kremser}
\affiliation{Walter Schottky Institut, Technische Universit\"at M\"unchen, 85748 Garching bei M\"unchen, Germany}
\author{Jonathan J. Finley}
\affiliation{Walter Schottky Institut, Technische Universit\"at M\"unchen, 85748 Garching bei M\"unchen, Germany}
\author{Kai M\"uller}
\affiliation{Walter Schottky Institut, Technische Universit\"at M\"unchen, 85748 Garching bei M\"unchen, Germany}
\author{Jelena Vu\v{c}kovi\'c}
\affiliation{E. L. Ginzton Laboratory, Stanford University, Stanford CA 94305, USA}

\date{\today}

\begin{abstract}
The Area Theorem states that when a short optical pulse drives a quantum two-level system, it undergoes Rabi oscillations in the probability of scattering a single photon. In this work, we investigate the breakdown of the Area Theorem as both the pulse length becomes non-negligible and for certain pulse areas. Using simple quantum trajectories, we provide an analytic approximation to the photon emission dynamics of a two-level system. Our model provides an intuitive way to understand re-excitation, which elucidates the mechanism behind the two-photon emission events that can spoil single-photon emission. We experimentally measure the emission statistics from a semiconductor quantum dot, acting as a two-level system, and show good agreement with our simple model for short pulses. Additionally, the model clearly explains our recent results [K. Fischer and L. Hanschke, et al., \textit{Nature Physics} (2017)] showing dominant two-photon emission from a two-level system for pulses with interaction areas equal to an even multiple of $\pi$.
\end{abstract}

\pacs{Valid PACS appear here}
\maketitle


\section{Introduction}

One of the most fundamental building blocks of quantum optics is the single discrete atomic transition, which at its simplest is modeled as a quantum two-level system \cite{Cohen-Tannoudji1992-uo}. This type of system has been behind fundamental discoveries such as photon anti-bunching \cite{Michler2000-dx,Kimble1977-rw}, Mollow triplets \cite{Flagg2009-zi,Mollow1969-dv}, and quantum interference of indistinguishable photons \cite{Santori2002-wi,Hong1987-lv}. After almost two decades of development in a solid-state environment, the quantum two-level system is now poised to serve the pivotal role of an on-demand single-photon source \cite{Michler2000-dx,Santori2002-wi,he2013demand,schneider2015single,yuan2002electrically,claudon2010highly,unsleber2016highly,schlehahn2016electrically,somaschi2016near,ding2016demand,michler}---by converting laser pulses with Poissonian counting statistics to single photons---for quantum networks \cite{Loredo2016-jz,Aharonovich2016-pq,o2009photonic}. More recently, multi-photon quantum state generators have generated strong interest as replacements for the single-photon source in many quantum applications \cite{Rundquist2014-kf,munoz2014emitters,afek2010high}. To this end, it was recently discovered that two-level systems may also generate pulses containing two photons \cite{fischer2017signatures}. In this article, we show a simple and intuitive model to better understand the temporal excitation dynamics of quantum two-level systems.

Consider an ideal two-level system \cite{Shore2011-fz}, with a ground state $\ket{g}$ and an excited state $\ket{e}$. Suppose the system is driven by an optical pulse starting at $t=0$, resonant with the $\ket{g}\leftrightarrow\ket{e}$ transition and where the rotating wave approximation holds. As a function of the integrated pulse area
\begin{equation}
A(t) = {\int_{0}^t \mathop{\textrm{d} t'} \mu\cdot E(t')/\hbar},\label{eq:area}
\end{equation}
where $E(t')$ is the envelope of the pulse's electric field and $\mu$ the system's electric dipole moment, the system undergoes coherent oscillations between its ground $\ket{g}$ and excited $\ket{e}$ states. If the system is initially prepared in the ground state, the state after the system-pulse interaction is given by
\begin{equation}
\ket{\psi_f(A)}=\sqrt{1-\textrm{P}_e(A(t))}\ket{g}+\textrm{e}^{-\textrm{i}\phi}\sqrt{\textrm{P}_e(A(t))}\ket{e},
\end{equation}
where $\phi$ is a phase set by the laser field and $\textrm{P}_e(A(t))$ is the area-dependent probability of exciting the two-level system. Examining $\textrm{P}_e(A(t))$ shows Rabi oscillations that are perfectly sinusoidal
\begin{equation}
\textrm{P}_e(A(t))=\textrm{sin}^2(A(t)/2),
\end{equation}
with the laser pulse capable of inducing an arbitrary number of rotations between $\ket{g}$ and $\ket{e}$. After the system has interacted with the entire pulse, i.e. in the limit of $t\rightarrow\infty$, the probability of remaining in the excited state is
\begin{equation}
\textrm{P}_e(A(\infty))=\textrm{sin}^2(A(\infty)/2).
\end{equation}
This statement is called the Area Theorem, in which Rabi oscillations occur as a function of the total interaction area of the pulse.  The majority of a pulse's area typically interacts within a time window $T$, referred to as the width of the pulse. We can state this mathematically as
\begin{equation}
A(T)\approx A(\infty).
\end{equation}

However, a realistic system is coupled to the outside world through its electric dipole and may spontaneously decay, which spoils the results of the Area Theorem \cite{Rodgers1987-hc,Zakrzewski1986-cw,fischer2017signatures,dada2016indistinguishable}. Previous works have shown its regime of validity occurs when the width of the pulse $ T $ is very short compared to the spontaneous decay time $\tau_e=1/\gamma$. That way, the system-pulse interaction occurs before any spontaneous emissions---we now explore this concept from a photon counting perspective \cite{carmichael2009open}.


\section{A theory for photon counting\label{sec:photcnt}}

Spontaneous emission of photons can be thought of as turning the system's wavefunction into a stochastic process, and realizations of this process are called trajectories (see Appendix A for a formal justification of the following counting procedure). At any time step, a photon emission may occur at a rate of $\gamma \text{P}_e$, resetting the wavefunction into its ground state and `restarting' the Area Theorem. The Area Theorem then again results in deterministic evolution of the system. This process occurs until the pulse has been fully interacted, after which the two-level system can only decay spontaneously. Then, we label a resulting trajectory by a set of emission times $t_1<\cdots<t_n$, and this trajectory occurs with the probability density
\begin{eqnarray}
p_n(t_1,\cdots,t_n).
\end{eqnarray}
Integrating these densities yields the total probability for $n$ photon emissions
\begin{eqnarray}
P_n=\int_0^\infty \mathop{\text{d}t_1}\int_{t_1}^\infty \mathop{\text{d}t_2}\cdots \int_{t_{n-1}}^\infty \mathop{\text{d}t_n}p_n(t_1,\cdots,t_n).
\end{eqnarray}
Because photon emissions are sequential, we will be able to more naturally construct inclusive probabilities that represent the probability for $n$ or more emissions
\begin{eqnarray}
F_n=P_n+P_{n+1}+P_{n+2}+\cdots.
\end{eqnarray}
They have associated densities as well $f_n(t_1,\cdots,t_n)$, which correspond to any emission sequence that has its first emissions times as $t_1<\cdots<t_n$, and implicitly include the probability of any subsequent emissions as well. They are similarly related to their parent probabilities as
\begin{eqnarray}
F_n=\int_0^\infty \mathop{\text{d}t_1}\int_{t_1}^\infty \mathop{\text{d}t_2}\cdots \int_{t_{n-1}}^\infty \mathop{\text{d}t_n} f_n(t_1,\cdots,t_n).\label{eq:inclusiveP}
\end{eqnarray}
If we can solve for $F_n$, they are trivially related to the exclusive emission probabilities
\begin{eqnarray}
P_n=F_n-F_{n+1}\label{eq:ItoE}
\end{eqnarray}
with the special case $P_0=1-F_1$.

We now outline the computing of each $f_n(t_1,\cdots,t_n)$ for a short pulse, beginning with $f_1(t_1)$---the density that a first emission occurs at time $t_1$. Consider the time period during the pulse $0<t_1<T$. The probability no photon was emitted before $t_1$ is given by
\begin{eqnarray}
\text{e}^{-\lambda}\approx \text{e}^{-\gamma t_1/2}\;\text{ if }0<t_1<T,\quad\quad\label{eq:approxp0}
\end{eqnarray}
because the emissions behave like a Poisson point process with variable rate parameter
\begin{eqnarray}
\lambda=\int_0^{t_1} \mathop{\text{d}t} \gamma \text{P}_e(A(t))\;\text{ if }0<t_1<T,
\end{eqnarray}
and the Area Theorem sets this variable rate (see Appendix A for more details). The approximation in Eq. \ref{eq:approxp0} is valid in the short pulse limit, where the \textit{average} excited state population dominates the probability no photon was emitted over an interval. Putting this together with the true instantaneous emission rate $\gamma \text{P}_e(A(t_1))$, the inclusive probability density is
\begin{equation}
f_1(t_1)\approx\gamma\,\textrm{sin}^2(\tfrac{A(t_1)}{2})\,\text{e}^{-\gamma t_1/2} \;\text{ if }0<t_1<T.\label{eq:density1}
\end{equation}
In the time period after the pulse $T<t_1$, the decay of the density is exponential since the system can only emit spontaneously, yielding
\begin{equation}
f_1(t_1)\approx\gamma\,\textrm{sin}^2(\tfrac{A(\infty)}{2})\,\text{e}^{-\gamma T/2}\text{e}^{-\gamma(t_1-T)} \;\text{ if }T< t_1.\label{eq:approx13}
\end{equation}
(We refer the reader to Appendix A for a formal derivation of this rate.) The associated inclusive probability is
\begin{eqnarray}
F_1&=&\gamma\int_0^T \mathop{\text{d}t_1} \textrm{sin}^2(\tfrac{A(t_1)}{2})\,\text{e}^{-\gamma t_1/2}+\hspace{10ex}\\
&&\hspace{-0ex}\gamma\,\textrm{sin}^2(\tfrac{A(\infty)}{2})\,\text{e}^{-\gamma T/2}\int_T^\infty \mathop{\text{d}t_1} \text{e}^{-\gamma(t_1-T)}\nonumber\\
&\approx& \gamma\,\text{e}^{-\gamma T/2}\int_0^T \mathop{\text{d}t_1} \textrm{sin}^2(\tfrac{A(t_1)}{2}) +\label{eq:F1}\\
&&\hspace{20ex}\textrm{sin}^2(\tfrac{A(\infty)}{2})\,\text{e}^{-\gamma T/2},\nonumber
\end{eqnarray}
where the approximation that
\begin{eqnarray}
\gamma\int_0^T \mathop{\text{d}t_1} \textrm{sin}^2(\tfrac{A(t_1)}{2})\,\text{e}^{-\gamma t_1/2}\approx&\\
&\hspace{-10ex}\gamma\,\text{e}^{-\gamma T/2}\int_0^T \mathop{\text{d}t_1} \textrm{sin}^2(\tfrac{A(t_1)}{2})\nonumber
\end{eqnarray}
holds for short pulses. One might be tempted to further approximate away the term with the integral in Eq. \ref{eq:F1} since it's $\mathcal{O}(\gamma T)$ and the second term looks $\mathcal{O}(1)$, however, Rabi oscillations can cause the second term to vanish when $A(\infty)=\{2\pi,4\pi,\cdots\}$. Then, the $\mathcal{O}(\gamma T)$ term dominates the emission density.

Next, we construct the inclusive probability density for a second emission at time $t_2$ following the first emission at time $t_1$. The first emission resets the system to its ground state, so the application of the Area Theorem restarts from $t_1$. This leads to the new factor $\gamma\,\text{P}_e(A(t_2)-A(t_1))$ in $f_2$, so the density of two emissions occurring during the pulse is given by
\begin{eqnarray}
f_2(t_1,t_2)\approx\gamma^2\,\textrm{sin}^2(\tfrac{A(t_2)-A(t_1)}{2})\,\textrm{sin}^2(\tfrac{A(t_1)}{2})\,\text{e}^{-\gamma t_2/2} &\nonumber\\
&\hspace{-15ex}\;\text{ if }t_2<T.\quad\quad\label{eq:mag0}
\end{eqnarray}
When the first emission occurs during the pulse and the second after, then
\begin{eqnarray}
f_2(t_1,t_2)\approx\gamma^2\,\textrm{sin}^2(\tfrac{A(\infty)-A(t_1)}{2})\,\textrm{sin}^2(\tfrac{A(t_1)}{2})\cdot &\\
&\hspace{-25ex}\text{e}^{-\gamma T/2}\text{e}^{-\gamma(t_2-T)}\;\text{ if }t_1<T<t_2.\quad\quad\label{eq:mag1}\nonumber
\end{eqnarray}
Because there is no energy after the pulse to re-excite the system following an emission,
\begin{eqnarray}
f_2(t_1,t_2)=0\;\text{ if }T<t_1<t_2.
\end{eqnarray}
The density from $t_2<T$ contributes $\mathcal{O}((\gamma T)^2)$ to $F_2$ but the density from $T<t_2$ contributes $\mathcal{O}(\gamma T)$, for \textit{any} total area $A(T)=A(\infty)$. Hence, in our short pulse approximation we take only the second interval
\begin{eqnarray}
\hspace{-5ex}F_2&\approx& \gamma\,\text{e}^{-\gamma T/2}\int_0^T \mathop{\text{d}t_1} \textrm{sin}^2(\tfrac{A(\infty)-A(t_1)}{2})\,\textrm{sin}^2(\tfrac{A(t_1)}{2}),\label{eq:F2}
\end{eqnarray}
where we have already integrated out $t_2$---that integral sums over simple exponential decay.

Each new emission during the pulse time resets the Area Theorem and contributes a factor of $\gamma\,\text{P}_e(A(t_n)-A(t_{n-1}))$ to $f_n$. Hence, the general solution for an emission sequence $0<t_1<\cdots<t_n$ with a short pulse is
\begin{widetext}
\begin{eqnarray}
\hspace{-4ex}f_n(t_1,\cdots,t_n)\approx&\nonumber\\&\hspace{-10ex}
\begin{cases}
\gamma^n\,\textrm{sin}^2(\frac{A(t_n)-A(t_{n-1})}{2})\cdots\textrm{sin}^2(\frac{A(t_2)-A(t_1)}{2})\,\textrm{sin}^2(\frac{A(t_1)}{2})\,\text{e}^{-\gamma t_n/2} &\text{ if }t_n<T\\
\gamma^n\,\textrm{sin}^2(\frac{A(\infty)-A(t_{n-1})}{2})\,\textrm{sin}^2(\frac{A(t_{n-1})-A(t_{n-2})}{2})\cdots\\
\hspace{30ex}\cdots\textrm{sin}^2(\frac{A(t_2)-A(t_1)}{2})\,\textrm{sin}^2(\frac{A(t_1)}{2})\,\text{e}^{-\gamma T/2}\text{e}^{-\gamma(t_n-T)} &\text{ if }t_{n-1}<T<t_n\\
0 & \text{ otherwise}\label{eq:fngeneral}
\end{cases}.
\end{eqnarray}
\end{widetext}
In the case where we integrate to obtain the counting probabilities (Eq. \ref{eq:inclusiveP}), the density from $t_n<T$ contributes $\mathcal{O}((\gamma T)^n)$ but the density from $T<t_n$ contributes $\mathcal{O}((\gamma T)^{n-1})$, for \textit{any} total area $A(T)=A(\infty)$. Then, we further approximate
\begin{widetext}
\begin{eqnarray}
F_n\approx\gamma^{n-1}\text{e}^{-\gamma T/2}
\int_0^T \mathop{\text{d}t_1}\int_{t_1}^T \mathop{\text{d}t_2}\cdots \int_{t_{n-2}}^T \mathop{\text{d}t_{n-1}}\,\textrm{sin}^2(\tfrac{A(\infty)-A(t_{n-1})}{2})\,\textrm{sin}^2(\tfrac{A(t_{n-1})-A(t_{n-2})}{2})\cdots&\nonumber\\
&\hspace{-20ex}\cdots\textrm{sin}^2(\frac{A(t_2)-A(t_1)}{2})\,\textrm{sin}^2(\frac{A(t_1)}{2}),\label{eq:FN}
\end{eqnarray}
\end{widetext}
where we have already integrated out $t_n$. This is equivalent to saying that we expect every photon emission to occur during the pulse except for the last one, which occurs afterwards. Because of the exponential factor in front, this set of inclusive probabilities is almost always properly normalized. We briefly note that in the limit of long pulses
\begin{eqnarray}
F_n\approx \frac{(\gamma T/2)^{n-1}}{(n-1)!}\text{e}^{-\gamma T/2},
\end{eqnarray}
which is Poisson distributed.

Back to short pulses: the inclusive probabilities die away with increasing number of $n$ emissions. Therefore, the exclusive probabilities $P_n$ are all $\mathcal{O}(( \gamma T)^{n-1})$ for $n>1$. As the pulse length increases, our model will deviate from the exact behavior since we neglected the cases where all $n$ photon emissions occur during the pulse. It will also deviate from ignoring the effect of spontaneous emission on the deterministic dynamics during the pulse (see Appendix A), which is justified in the strong driving or short pulse limit.


\section{Breakdown with increasing pulse length}

\begin{figure*}
  \includegraphics[width=\textwidth]{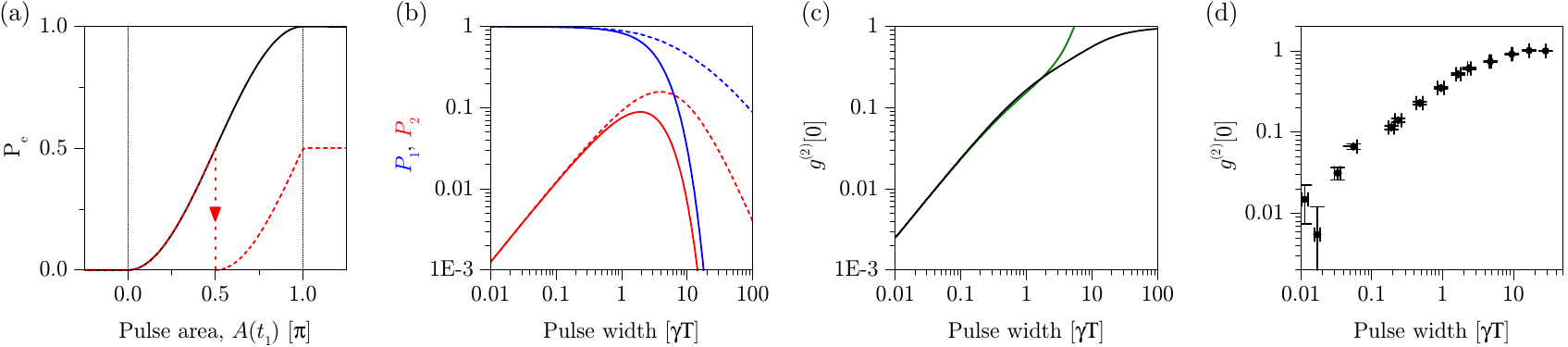}
  \caption{Re-excitation dynamics for two-level system under interaction with a $\pi$-pulse, i.e. $A(\infty)=\pi$. (a) Two example quantum trajectories, with no photon emission (solid black) and one photon emission (dashed red) during the system-pulse interaction. Arrow indicates time of photon emission. (b) Probability of one photon emission $P_1$: simple model (solid blue) and numerically exact (dashed blue). Probability of two photon emissions $P_2$: simple model (solid red) and numerically exact (dashed red). Both under excitation by a square pulse. (c) Measured degree of second-order coherence $g^{(2)}[0]$ under excitation by a square pulse, simulated with the quantum regression theorem (black) and estimated from our simple analytic approach (green). (d) Experimentally measured degree of second-order coherence, obtained when resonantly exciting a two-level system made up of excitonic states from a deterministically charged quantum dot.}
  \label{figure:1}
\end{figure*}

Although remarkably simple, our trajectory formalism allows for analytic exploration of an important phenomenon in solid-state single-photon sources---many single-photon sources behave as nearly ideal quantum two-level systems \cite{schneider2015single,Wang2016-dq,Kuhlmann2015-hh}. These systems operate by sending a short pulse of area $A(\infty)=\pi$ to excite the system to $\ket{e}$ with almost unity probability, which results in emission of a single-photon wavepacket with high probability \cite{Fischer2016-pl}. The target single-photon wavepackets may be used in quantum information systems in the future, however, some of these systems have extremely stringent requirements against multiphoton errors \cite{azuma2015all,kok2007linear}. Thus, our proposed model for understanding multiphoton errors, as due to re-excitation in a two-level system, can be used to help identify useful regimes of operation.

To explore this concept further, we suppose the excitation of a two-level system by a short pulse of area $A(\infty)=\pi$. The dynamics of this process as a function of interacted pulse area are depicted in Fig.~\ref{figure:1}a. A standard quantum trajectory \cite{carmichael2009open} with no re-excitation is shown as the black curve, where the single-photon emission would occur some time long after the system-pulse interaction. Because the $n$-photon generation probability scales as $( \gamma T)^{n-1}$, we only need to consider the error due to a single re-excitation. The red dashed curve depicts a trajectory with an emission occurring after $A(t_1)=\pi/2$ has been interacted. Similar to the trajectory with a single emission, the re-excited system is most likely to emit its second photon long after the system-pulse interaction has occurred.

By integrating over all possible trajectories, we can compute the signal rate $P_1(A(\infty)=\pi)$ and error rate $P_2(A(\infty)=\pi)$, which was only extracted experimentally \cite{dada2016indistinguishable} and numerically \cite{Fischer2016-pl} thus far. We now assume a specific form of $A(t)$ corresponding to a square pulse, so 
\begin{eqnarray}
A(t)=
\begin{cases}
t\frac{A(\infty)}{T}&\;\text{ if } 0<t<T\\
A(\infty)&\;\text{ if } T<t
\end{cases}.
\end{eqnarray}
From Eqs. \ref{eq:F1}, \ref{eq:F2}, and \ref{eq:FN}, we identify the short pulse approximations to the inclusive emission probabilities and explicitly write them for the square pulse
\begin{eqnarray}
F_1(A(\infty)=\pi)&\approx&\text{e}^{-\gamma T/2} \left(1+\frac{\gamma T}{2}\right)\\
F_2(A(\infty)=\pi)&\approx&\text{e}^{-\gamma T/2} \frac{\gamma T}{8 } \nonumber\\
F_3(A(\infty)=\pi)&\approx& \text{e}^{-\gamma T/2} \frac{-24 + \pi^2 + 4 \left(-6 + \pi^2\right) \left(\gamma T\right)^2 }{64 \pi^2}\nonumber.
\end{eqnarray}
The differences between inclusive probabilities give the exclusive probabilities (Eq. \ref{eq:ItoE}) as $P_1=F_1-F_2$ and $P_2=F_2-F_3$, with
\begin{eqnarray}
P_1(A(\infty)=\pi)&\approx&\text{e}^{-\gamma T/2} \left(1+\frac{3\gamma T}{8}\right)\label{eq:P1P2pipulse}\\
P_2(A(\infty)=\pi)&\approx&\text{e}^{-\gamma T/2} \left(\frac{\gamma T}{8 } + \left(\gamma T\right)^2\left(\frac{3 }{4\pi^2}-\frac{5}{64}\right)\right) \nonumber
\end{eqnarray}
shown as the solid curves in Fig. \ref{figure:1}b. We also computed the exclusive probabilities numerically without approximation for the pulse (dashed curves). Notably, our approximations hold fairly well until $\gamma T$ is $\mathcal{O}(1)$.

These probabilities can then be used to estimate the experimentally measurable metric of the quality of a single-photon source \cite{Fischer2016-pl}, the degree of second-order coherence
\begin{equation}
g^{(2)}[0]=\frac{\sum_k k(k-1) P_k }{(E[n])^2}
\end{equation}
with the expected photon number
\begin{equation}
\text{E}[n]=\sum_k k P_k.
\end{equation}
As the pulse-width decreases $g^{(2)}[0]$ approaches from $1\rightarrow0$, shown as the black curve in Fig.~\ref{figure:1}c, indicating an increase in the quality of the single-photon nature of the wavepacket (see Appendix B for a discussion on the temporally resolved second-order coherence). The degree of second-order coherence was computed as in Ref. \cite{Fischer2016-pl} with the Quantum Optics Toolbox in Python (QuTiP) \cite{Johansson2013-cv}. For short pulses, the system acts as a good single-photon source and 
\begin{eqnarray}
g^{(2)}[0]&\approx&\frac{2 P_2 }{(P_1+2 P_2)^2},
\end{eqnarray}
which we can calculate from our analytical expressions in Eq. \ref{eq:P1P2pipulse} (solid green curve).

Any ideal Rabi oscillation of equivalent $A(\infty)$ is isomorphic by a nonlinear coordinate transformation in time because the Area theorem depends on $A(t)$ (e.g. in Fig. \ref{figure:1}a the $x$-axis is $A(t)$). However, the precise form of the pulse can change $P_2$ for a given $T$ due to integration over the time variable. For extremely short pulses the exponential factors in $P_n$ are irrelevant, and for example, a Gaussian pulse yields $P_2\approx 0.2188 \,\gamma T$ while a square pulse yields $P_2\approx\gamma T/8$. (We note here that we considered an incident Gaussian pulse where $ T $ is the width in energy, by convention in the field of single-photon sources.)

Our approximation works quite well until $ \gamma T \approx1$, where the analytic model does not correctly capture the saturation behavior of the second-order coherence. For long pulse lengths, the statistics tend towards Poissonian due to the possibility of many randomly distributed photon emissions occurring.

\begin{figure*}
  \includegraphics[width=\textwidth]{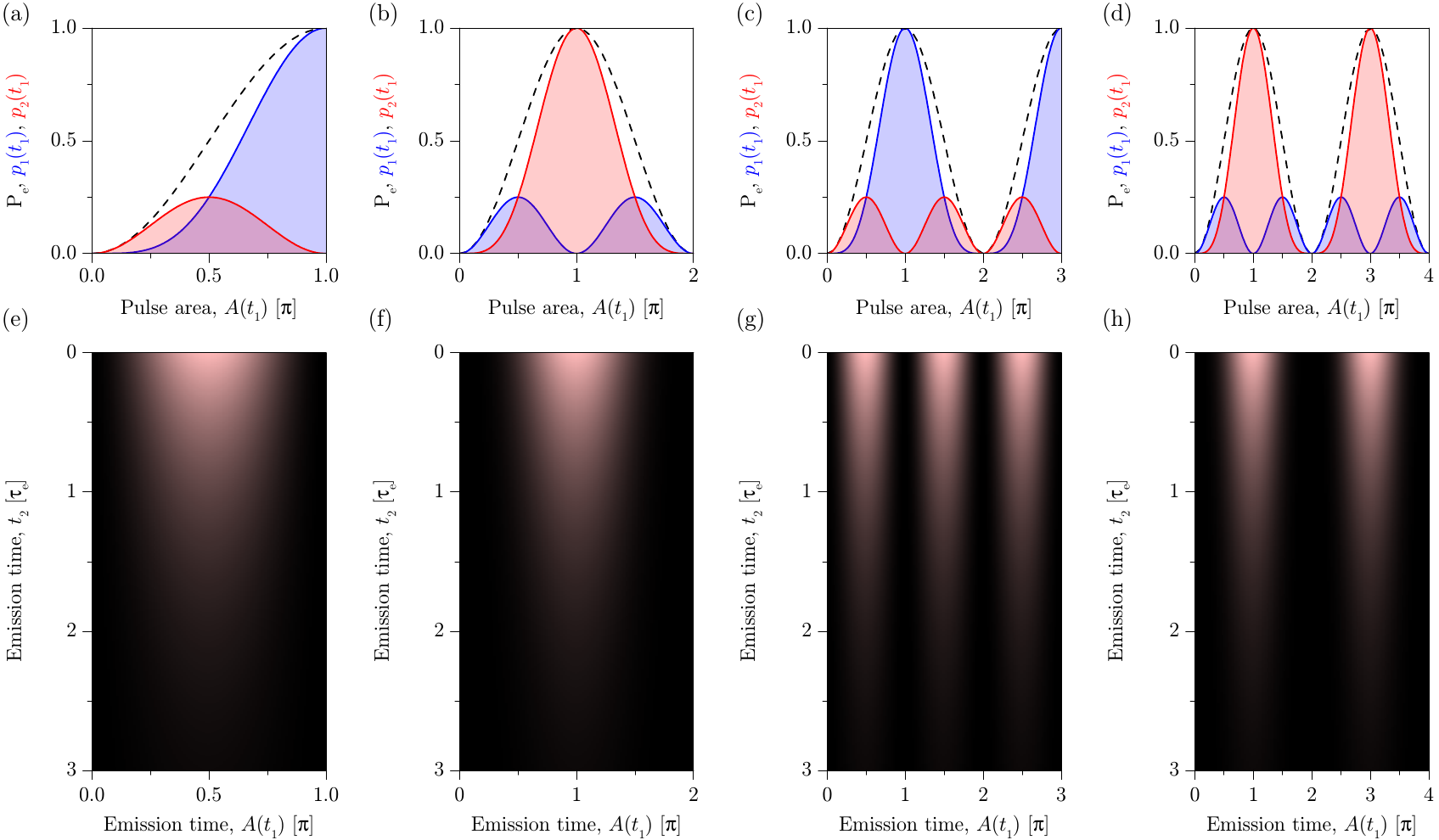}
  \caption{Re-excitation dynamics for a two-level system under interaction with extremely short $m\pi$-pulses, i.e. $A(\infty)=m\pi$ for $m\in\{1,2,3,4\}$. Here, the precise pulse shape is not important because the exponential factors in Eq. \ref{eq:FN} are all nearly unity for extremely short pulses, so we use the pulse-independent function $A(t_1)$ as our x-axis. (a)-(d) Probability of system being in the excited state $\textrm{P}_\textrm{e}$ if no emissions occur, and probability density $p_1(t_1)$ for a single photon emission at time $t_1$ and $p_2(t_1)$ for a pair of photon emissions to begin at time $t_1$. (e)-(h) Fully time-resolved probability density $p_2(t_1, t_2)$ for a pair of photon emissions, assuming the system-pulse interaction time is very short compared to the excited state lifetime, i.e. $ \gamma T \ll1$.}
  \label{figure:2}
\end{figure*}

To verify that our simple model has real predictive power for short pulses, we experimentally measured the degree of second-order coherence $g^{(2)}[0]$ of pulses scattered by a resonantly driven two-level system (Fig.~\ref{figure:1}d). We explored a much broader range of pulse widths compared to previous data, showing a complete series from a few ps to 10's of ns, which allows us to explore how our analytic model deviates from experiment for long pulses as well. Our two-level system of choice was the single electron to trion transition of a charged InAs/GaAs quantum dot (see Supplementary material for data and details on the experiments). Good agreement can be seen between the full quantum-optical model and data, with only small differences that result from experimental inaccuracies (see Supplementary material). Meanwhile, the analytic model gives good agreement until $ \gamma T \approx 1$.

In summary, we have quantified how the Area Theorem breaks down as a function of pulse length. This discussion can provide an important limit on the achievable source error rates for a given ratio of pulse length to spontaneous emission lifetime, an important addition to previous studies on the phonon-induced limitations for emission of indistinguishable photons from solid-state devices \cite{iles2017phonon,kaer2013microscopic,kaer2013role,thoma2016exploring}. We expect new directions in research on solid-state sources to involve understanding the interplay of pulsed excitation with phonon dynamics \cite{gustin2017influence,muller2016self,muller2015ultrafast}. For short pulses, we believe the phonon dynamics work to incoherently populate (or re-excite) the system---most likely with the same type of linear dependence as the coherent excitation mechanism we have identified \cite{fischer2017signatures}.


\section{Breakdown for even pi areas}

As recently discovered \cite{fischer2017signatures}, the Area Theorem breaks down when $A(\infty)=2m\pi$ and $m\in\{1,2,3,...\}$. We now explore this point with our simple model. To help understand this breakdown we introduce the new probability densities
\begin{eqnarray}
p_1(t_1)&= &f_1(t_1) - \int_{t_1}^\infty \mathop{\text{d} t_2} f_2(t_1,t_2),
\end{eqnarray}
which represents the exclusive and marginal probability density for a single photon emission, and
\begin{eqnarray}
p_2(t_1)&= &\int_{t_1}^\infty \mathop{\text{d} t_2}\left(f_2(t_1,t_2) - \int_{t_2}^\infty \mathop{\text{d} t_3} f_2(t_1,t_2,t_3)\right)\quad\quad,
\end{eqnarray}
which represents the exclusive and marginal probability density for a single photon emission to be the start of a two-photon emission sequence. In the limit of very short pulses, where the exponential factors do not matter and we keep only terms of comparable order in $\gamma T$,
\begin{eqnarray}
p_1(t_1)&\approx &\gamma \left(\textrm{sin}^2(\tfrac{A(t_1)}{2}) - \textrm{sin}^2(\tfrac{A(\infty)-A(t_1)}{2})\,\textrm{sin}^2(\tfrac{A(t_1)}{2})\right)\nonumber\\
&&\hspace{30ex}\;\;\;\;\text{ if } t_1<T\nonumber\\
&=&\gamma\,\textrm{sin}^2(\tfrac{A(t_1)}{2})\,\textrm{cos}^2(\tfrac{A(\infty)-A(t_1)}{2})\;\text{ if } t_1<T
\end{eqnarray}
and
\begin{eqnarray}
p_2(t_1)&\approx &\int_{t_1}^\infty \mathop{\text{d} t_2} f_2(t_1,t_2)\\
&=&\gamma T\, \textrm{sin}^2(\tfrac{A(\infty)-A(t_1)}{2})\,\textrm{sin}^2(\tfrac{A(t_1)}{2}).
\end{eqnarray}

Consider the probability densities for a $\pi$ pulse (Fig.~\ref{figure:2}a), plotted independent of the pulse shape by using $A(t_1)$ as the $x$-axis. The trajectory if no photon emission occurs is again shown as a function of interacted pulse area; however, we now additionally show $p_1(t_1)$ and $p_2(t_1)$. Integrals of the densities give the probabilities for one or two photon emissions to occur, respectively, with the only or first emission to occur during the system-pulse interaction time. The shaded regions hence depict total probabilities for exclusive types of emission events (though the exact values are determined by the specific pulse shape). By considering the fully time-resolved probability density for two photon emissions, we comment that $p_2(t_1, t_2)$ is approximately separable for short pulses as $p_2(t_1, t_2)\approx p_2(t_1)\, \textrm{e}^{-\gamma t_2}$, which is plotted for a $\pi$ pulse in (Fig.~\ref{figure:2}e). Additionally, consider two other scenarios:
\begin{enumerate}
\item The situation is dramatically different when the pulse is long compared to the emission lifetime and multiple re-excitation events can occur (see Appendix B).
\item The results for $A(\infty)=\pi$ are similar to all pulses with odd-$\pi$ areas. For example, examine the densities for $\pi$ (Fig.~\ref{figure:2}a,e) and $3\pi$ (Fig.~\ref{figure:2}c,g) pulse areas: they are comparable whereby the densities from $A(\infty)=\pi$ have been tessellated three times for $A(\infty)=3\pi$.
\end{enumerate}

On the other hand, the situation for $p_1(t_1)$ is radically different for a $2\pi$ pulse (Fig.~\ref{figure:2}b,f). Importantly, the Area Theorem requires the excited-state population to return to zero after the system-pulse interaction, as shown by the dashed black line. Hence, both $P_2$ \textit{and} $P_1$ are of order $ \gamma T$. This procedure yields $P_2/P_1\approx2.33$ for a Gaussian pulse and $P_2/P_1=3$ for a square pulse, with the remarkable result that $P_2 > P_1$. Interestingly, $P_2$ is emphasized because the most likely initial emission occurs when roughly $\pi$ of the area has been absorbed, forcing a re-excitation with almost unity probability by the remaining $\pi$ area. In fact, our results in this regime are nearly identical for all pulses of even-$\pi$ area (see Fig.~\ref{figure:2}d,h for $A(\infty)=4\pi$, where the probability densities for $A(\infty)=2\pi$ have just been copied a second time). Nearly constant ratios of $P_2/P_1$ for all even-$\pi$ areas occur because both $p_1(t_1)$ and $p_2(t_1)$ are periodic with area $2\pi$. The most important consequence of the periodicity is that $P_2> P_1$ for even arbitrarily short pulses, showing a dramatic result in the breakdown of the Area Theorem when $A(\infty)=2m\pi$.

We comment on the pulse-length dependence of $P_1$ and $P_2$ for $2\pi$ pulses in Appendix C. It turns out that even for quite long pulses, two-photon emission is strongly dominant all the way up to $\gamma T\approx1$. This bodes well for future experiments and its potential use as a two-photon source, and we expect that further pulse-shape optimization could lead to even stronger two-photon emission. We discuss the extension to three-photon emission densities in Appendix D, in order to explore why the two-photon emission can dominate for such long pulse lengths.


\section{Comparison to realistic rabi oscillations}

\begin{figure}[]
  \includegraphics[width=\columnwidth]{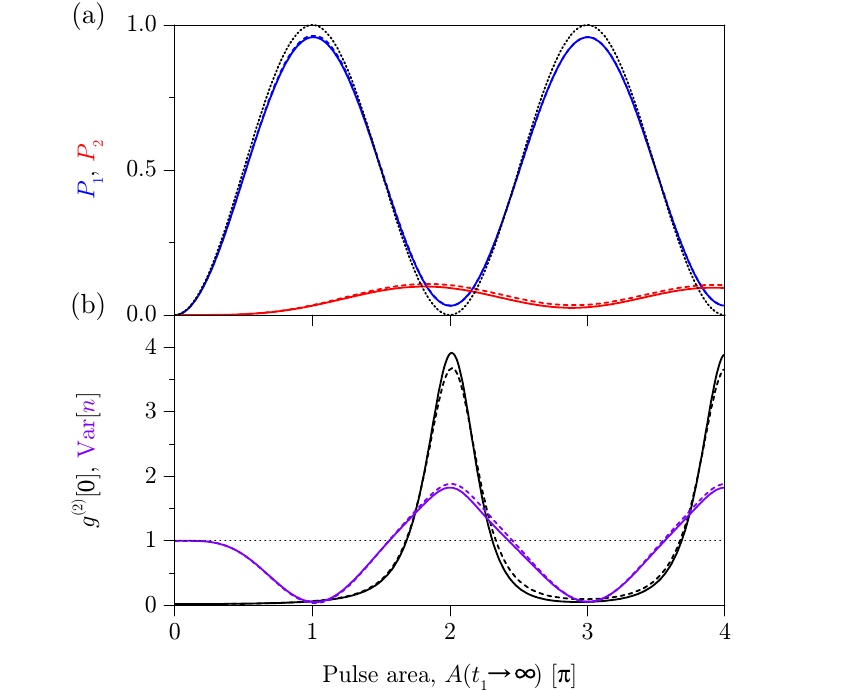}
  \caption{Photon emission statistics of a two-level system under excitation by a square pulse of $T=0.3/\gamma$. (a) Probabilities for $P_1$ single and $P_2$ pairs of photon emissions. Black curve shows $P_1$ for an ideal Rabi oscillation, while blue and red dashed curves indicate simulated values using a full quantum trajectory calculation that includes spontaneous emission. Solid red and solid blue curves show estimates based on our simple analytic approach. (b) Exact simulations for the variance relative to a coherent state Var[$n$] and for $g^{(2)}[0]$ as dashed curves, while solid curves show estimates based on our simple analytic approach. Dotted black line shows coherence statistics of the incident laser pulse.}
  \label{figure:3}
\end{figure}

In the final section, we compare the results generated with our simple model to those from a full quantum mechanical simulation \cite{Fischer2016-pl,fischer2017signatures}. Here, we show the full inclusive probability densities calculated with our analytical model for a square pulse of width $T$:
\begin{widetext}
\begin{eqnarray}
F_1&\approx&\text{e}^{-\gamma T/2} \left(\text{sin}^2(\frac{A(\infty)}{2}) + \frac{\gamma T}{2 } \left(1 - \frac{\text{sin}(A(\infty))}{A(\infty)}\right)\right)\\
F_2&\approx&\text{e}^{-\gamma T/2} \frac{\gamma T}{8 } \left(2 + \text{cos}(A(\infty)) - 3 \frac{\text{sin}(A(\infty))}{A(\infty)}\right)\\
F_3&\approx& \text{e}^{-\gamma T/2} \frac{ (\gamma T)^2}{64 \,A(\infty)^2} \left(4 \left(-6 + A(\infty)^2\right) - \left(-24 + A(\infty)^2\right) \text{cos}(A(\infty)) + 9 A(\infty) \,\text{sin}(A(\infty))\right).
\end{eqnarray}
\end{widetext}
Again, the differences between the inclusive probabilities give the exclusive probabilities $P_1=F_1-F_2$ and $P_2=F_2-F_3$. We consider a typical pulse width of $T=0.3/\gamma$ and look at the Rabi oscillations in emitted photon number (Fig.~\ref{figure:3}a) and photon statistics (Fig.~\ref{figure:3}b) versus interacted pulse area. We choose this pulse length because the deviations between our analytic model and the numerically exact results become noticeable (for additional quantitative comparisons of this deviation at important points consider Figs. \ref{figure:1}b and \ref{figure:A3}).

Now, we discuss the Rabi oscillations in more detail. The ideal probabilities to scatter a single photon $P_1$, if spontaneous emission is all but ignored, are shown---these are the well-known sinusoidal Rabi oscillations predicted by the Area Theorem (dashed black curve). When the effects of spontaneous emission are included, re-excitation causes $P_1$ to deviate from perfect sinusoidal behavior (blue) and $P_2$ to become important (red). On this plot scale, $P_{n>2}$ are negligible. As we have already discussed, it is clear that for the even-$\pi$ pulses $P_2>P_1$.

We can also consider the photon statistics of the emission $g^{(2)}[0]$ and the variance of the emission relative to the variance of a coherent state
\begin{equation}
\textrm{Var}[n]=\frac{\sum_k \left(k^2 - (\textrm{E}[n])^2\right) P_k }{\textrm{E}[n]}.
\end{equation}
The relative photon-number variance is lowest around odd multiples of $\pi$, indicating the emission of a highly pure single-photon state. Around even-multiples of $\pi$, however, the emission is highly super-Poissonian because two-photon emission is emphasized, i.e. $P_2\gg P_1$. Thus the second-order coherence bunches, $g^{(2)}[0] > 1$, and the relative photon number variance peaks. All of these values can accurately be estimated from our simple model (solid curves) which agree very well with the full simulated values (dashed curves). This agreement furthers our assertion that the probability of three emissions $P_3$ is negligible because it is of order $( \gamma T)^{2}$. From our discussion, we clearly see the Area Theorem alone is incapable of correctly capturing the dynamics and photon statistics especially around $A(\infty)=2m\pi$. The failure holds even for arbitrarily short pulse areas, where the second-order coherence diverges resulting in a singularity.


\section{Outlook}

In this paper, we showed how the Area Theorem breaks down. We found an intuitive picture for understanding the photon emission dynamics from a two-level system and showed that a two-level system can mostly emit photon pairs for quite long pulses. In the future, we expect that further optimization of the pulse shape should allow for more pure two-photon emission. Finally, we believe it will be interesting to explore the separability of the two-photon pulses to understand the application of two-level systems as photon pair sources.


\section*{ACKNOWLEDGMENTS}

The authors thank Rahul Trivedi and Vinay Ramasesh for helpful discussions.

We gratefully acknowledge financial support from the National Science Foundation (Division of Materials Research---Grant. No. 1503759) and the Bavaria California Technology Center (BaCaTeC). KAF acknowledges support from the Lu Stanford Graduate Fellowship and the National Defense Science and Engineering Graduate Fellowship. LH acknowledges the DFG and BMBF for financial support via the Nanosystems Initiative Munich and Q.COM. MK acknowledges the International Max Planck Research School---Quantum Science and Technology---for financial support. KM acknowledges support from the Bavarian Academy of Sciences and Humanities.


\appendix

\section*{APPENDIX A: PHOTON COUNTING WITH THE STOCHASTIC SCHR{\"O}DINGER EQUATION}

Consider a more formal description of our two-level system driven by a short optical pulse, where we look at the evolution of the system's state vector
\begin{equation}
\ket{\psi(t)} = \psi_g(t)\ket{g}+\psi_e(t)\ket{e}.
\end{equation}
The optical driving occurs through the system's dipole operator $\sigma=\ket{g}\bra{e}$, resonant with the $\ket{g}\leftrightarrow\ket{e}$ transition, and where the rotating wave approximation holds. If the system-pulse interaction begins at $t=0$, then the Schr{\"o}dinger evolution is given by (using $\hbar=1$ and a rotating frame transformation) \cite{steck2007quantum,breuer2002theory}
\begin{equation}
\frac{\text{d}}{\text{d} t}\ket{\psi(t)} =-\text{i}\frac{\mu\cdot E(t)}{2}\left(\text{i}\sigma-\text{i}\sigma^\dagger\right)\ket{\psi(t)}\label{eq:appendix1},
\end{equation}
with the interacted pulse envelope from Eq. \ref{eq:area} and taking $\mu\cdot E(t)$ as real (our choice of phase is to provide real solutions). The system undergoes coherent Rabi oscillations between its ground and excited states: if the system is initially prepared in $\ket{g}$ then the solution to Eq. \ref{eq:appendix1} is
\begin{equation}
\ket{\psi(t)} =\cos{(A(t)/2)}\ket{g}+\sin{(A(t)/2)}\ket{e}.
\end{equation}
This is again, a statement of the Area Theorem.

However, the system also may spontaneously emit photons. Both spontaneous emissions and Rabi oscillations can be captured by the stochastic Schr{\"o}dinger equation \cite{steck2007quantum} 
\begin{eqnarray}
\text{d}\ket{\psi(t)} =-\text{i}\frac{\mu\cdot E(t)}{2}\left(\text{i}\sigma-\text{i}\sigma^\dagger\right)\ket{\psi(t)}\text{d}t + \nonumber\\
\frac{\gamma}{2}\left(\langle\sigma^\dagger\sigma\rangle(t)-\sigma^\dagger\sigma\right)\ket{\psi(t)}\text{d}t+\nonumber\\
\left(\frac{\sigma}{\sqrt{\langle\sigma^\dagger\sigma\rangle(t)}}-1\right)\ket{\psi(t)}\text{d}N(t,\psi).\label{eq:SME}
\end{eqnarray}
Importantly, this equation represents something called a quantum trajectory, which is a \textit{realization} of the stochastic process $\ket{\Psi(t)}$. The increment $\text{d}N(t,\psi)$ represents a Poisson process that `triggers' photon emission at random times. Its mean is 
\begin{eqnarray}
\langle\text{d}N(t,\psi)\rangle=\gamma \langle\psi(t)|\sigma^\dagger \sigma| \psi(t)\rangle\text{d}t \label{eq:ratee},
\end{eqnarray}
which is the instantaneous photon emission probability conditioned on the particular trajectory. The evolution of a trajectory $\ket{\psi(t)}$ is uniquely identified by a vector of random emission times $\vec{\tau}_n(t)=\{t_1,\cdots,t_n\}$ with $0<t_1<\cdots<t_n<t$, called the measurement record, and the total number of emissions $n$ is random. This vector implies that for these times $dN(\tau_i\in\vec{\tau}(t),\psi)=1$ and zero otherwise. These trajectories occur with probability $P(\vec{\tau}_n(t))$. We are interested in calculating the inclusive probability density of a given trajectory up to the $n$-th photon emission, so we want 
\begin{eqnarray}
f_n(t_1,\cdots,t_n)=P(\vec{\tau}_n(t_n)).
\end{eqnarray}
This density is inclusive because it does not specify the evolution of the trajectory after time $t_n$. We will now use this form of Eq. \ref{eq:SME} in a nonstandard way: as opposed to sampling the probability density for photon emissions by evolving many random trajectories and building a histogram of emission times, we directly compute their probability densities.

For short pulses, we can break up the stochastic evolution into two periods. First, the behavior during the system-pulse interaction $0<t<T$:
\begin{eqnarray}
\text{d}\ket{\psi(t)} \approx-\text{i}\frac{\mu\cdot E(t)}{2}\left(\text{i}\sigma-\text{i}\sigma^\dagger\right)\ket{\psi(t)}\text{d}t + \nonumber\\
\left(\frac{\sigma}{\sqrt{\langle\sigma^\dagger\sigma\rangle(t)}}-1\right)\ket{\psi(t)}\text{d}N(t,\psi)\nonumber\\
\text{ if }0<t<T,\label{eq:stochasticarea}
\end{eqnarray}
when $\gamma T\ll 1$. The first term again causes the system to Rabi oscillate like in the Area Theorem, and the second term collapses the wavefunction into the ground state through photon emission at random times. Each time an emission occurs $dN(t,\psi)=1$, and it resets the Area Theorem. For example, let's use this equation to calculate the probability density of emission $f_1(t_1)$ during the pulse. Specifically, we consider the evolution of $\ket{\psi(t)}$ until the first emission at time $t_1$, with the initial condition $\ket{\psi(0)}=\ket{g}$. Then, $dN(t<t_1,\psi)=0$ in Eq. \ref{eq:stochasticarea} and the evolution is deterministic
\begin{eqnarray}
\ket{\psi(t)}=\cos{(A(t)/2)}\ket{g}+\sin{(A(t)/2)}\ket{e}\nonumber\\
\;\text{ if }0<t<t_1<T.
\end{eqnarray}
Hence, the realization of the excited state probability is
\begin{eqnarray}
\text{P}_e(t)=\text{sin}^2(A(t)/2)\;\text{ if }0<t<t_1<T.
\end{eqnarray}
The rate of photon emission is given by Eq. \ref{eq:ratee}, i.e. $\langle\text{d}N(t<t_1,\psi)\rangle/\mathop{\text{d}t}=\gamma \,\text{P}_e(t)$. To arrive at the emission density, we also need to count the probability no photon was emitted before $t_1$. For a Poisson process this is given by $\text{e}^{-\lambda}$, and we have a variable rate parameter from Eq. \ref{eq:ratee} so
\begin{eqnarray}
\lambda(t_1,0)&=&\int_0^{t_1} \mathop{\text{d}t} \frac{\langle\text{d}N(t,\psi)\rangle}{\mathop{\text{d}t}}\\
&=&\int_0^{t_1} \mathop{\text{d}t} \gamma \text{P}_e(A(t))\;\text{ if }0<t_1<T.
\end{eqnarray}
Putting these all together
\begin{equation}
f_1(t_1)=\gamma\,\textrm{sin}^2(\tfrac{A(t_1)}{2})\,\text{e}^{-\lambda(t_1,0)} \;\text{ if }0<t_1<T,
\end{equation}
which in the short pulse limit is Eq. \ref{eq:density1}.

Second, consider the behavior after the system-pulse interaction $T<t$. The pulse no longer drives the system so
\begin{eqnarray}
\text{d}\ket{\psi(t)} \approx\frac{\gamma}{2}\left(\langle\sigma^\dagger\sigma\rangle(t)-\sigma^\dagger\sigma\right)\ket{\psi(t)}\text{d}t+\nonumber\\
\left(\frac{\sigma}{\sqrt{\langle\sigma^\dagger\sigma\rangle(t)}}-1\right)\ket{\psi(t)}\text{d}N(t,\psi)\nonumber\\
\text{ if }T<t.
\end{eqnarray}
If there were no emissions before $T$, then the initial condition of any trajectory is
\begin{eqnarray}
\ket{\psi(T)}=\cos{(A(T)/2)}\ket{g}+\sin{(A(T)/2)}\ket{e}.
\end{eqnarray}
Like before, suppose there are no emissions until time $t_1>T$, so $dN(t<t_1,\psi)=0$ and hence
\begin{eqnarray}
\ket{\psi(t)}=\cos{(\tfrac{A(T)}{2})}\ket{g}+\nonumber&\\
&\hspace{-20ex} \left( 1+\left(1/\text{sin}^2(\tfrac{A(T)}{2})-1\right)\text{e}^{\gamma (t-T)} \right)^{-1/2}\ket{e}  \nonumber\\
&\;\text{ if }T<t<t_1.\quad
\end{eqnarray}
Then, the rate of photon emission is given by
\begin{eqnarray}
\frac{\langle\text{d}N(T<t<t_1,\psi)\rangle}{\text{d}t}&=&\frac{\gamma\, \text{sin}^2(\tfrac{A(T)}{2}) \text{e}^{\gamma (t-T)}}{1 + \text{sin}^2(\tfrac{A(T)}{2}) \left(\text{e}^{\gamma (t-T)}-1\right)},\nonumber
\end{eqnarray}
and the probability for no emission to occur on this interval is
\begin{eqnarray}
\text{e}^{-\lambda(t_1,T)}=1 + \text{sin}^2(\tfrac{A(T)}{2}) \left(\text{e}^{\gamma (t_1-T)}-1\right).
\end{eqnarray}
Putting this together with the rate and the probability no emission occurred during $0<t<T$, the density is
\begin{eqnarray}
f_1(t_1)&=&\gamma\,\text{sin}^2(\tfrac{A(T)}{2})\,\text{e}^{-\gamma(t_1-T)}\,\text{e}^{-\lambda(T,0)} \\
&&\hspace{24ex}\text{ if }T<t_1.\nonumber
\end{eqnarray}
The decay is exponential and reduces to Eq. \ref{eq:approx13} in the short pulse limit.

By defining 
\begin{eqnarray}
\tilde{H}(t,\psi)=\frac{\mu\cdot E(t)}{2}\left(\text{i}\sigma-\text{i}\sigma^\dagger\right)+\text{i}\frac{\gamma}{2}\left(\langle\sigma^\dagger\sigma\rangle(t)-\sigma^\dagger\sigma\right),\nonumber
\end{eqnarray}
we can formally write down a solution for the inclusive probability density for the first emission occurring at time $t_1$, and an arbitrary pulse shape, as
\begin{eqnarray}
f_1(t_1)=R(t_1,0)\text{e}^{-\lambda(t_1,0)}
\end{eqnarray}
and
\begin{eqnarray}
R(t_2,t_1)=\gamma\,|\bra{e}\mathcal{T}\text{e}^{-\text{i}\int_{t_1}^{t_2}\mathop{\text{d}t}\tilde{H}(t,\psi)}\ket{g}|^2,
\end{eqnarray}
where $\mathcal{T}$ indicates time-ordering of the exponential. This exponential should be understood as a shorthand to integrate Eq. \ref{eq:SME} with $\text{d}N=0$. This process can be repeated sequentially---enumerating trajectories and their paths to get the inclusive probability densities. Generally for any number of emissions
\begin{widetext}
\begin{eqnarray}
f_n(t_1,\cdots,t_n)=R(t_n,t_{n-1})\text{e}^{-\lambda(t_{n},t_{n-1})}\cdots  R(t_2,t_1)\text{e}^{-\lambda(t_2,t_1)}  R(t_1,0)\text{e}^{-\lambda(t_1,0)}.
\end{eqnarray}
\end{widetext}


\vspace{20pt}
\section*{APPENDIX B: SECOND-ORDER COHERENCE WHEN DRIVEN BY A LONG PULSE}

\begin{figure}[b]
  \includegraphics[width=\columnwidth]{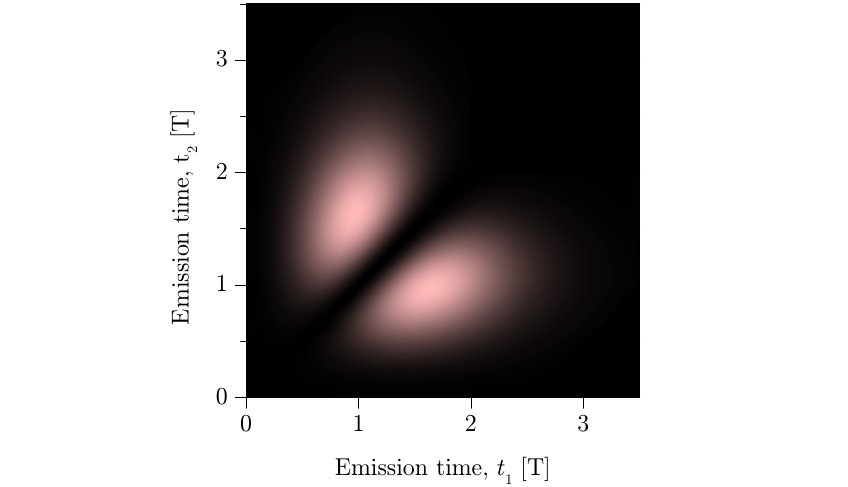}
  \caption{Second-order coherence $G^{(2)}(t_1, t_2)$ of emission from a two-level system under excitation by a long pulse ($ T =3.3/\gamma$) with area $A(\infty)=\pi$.}
  \label{figure:A1}
\end{figure}

In this appendix, we consider the temporally-resolved second-order coherence of the emission from a two-level system when it is driven by a long pulse ($ \gamma T \gg1$) with $\pi$ area. For short pulses, the second-order coherence is given by $G^{(2)}(t_1, t_2)=2\, p_2(t_1, t_2)$ for $t_2 > t_1$. Because we do not care if the first detection was at time $t_1$ or $t_2$, $G^{(2)}(t_1, t_2)$ is symmetric with exchange of these time indices. Thus, short pulses yield two thin slivers of second-order coherence that have dimensions $ T \times1/\gamma$ and $1/\gamma\times T $. However, for long pulses $G^{(2)}(t_1, t_2)$ is an inclusive \textit{moment} of the probability densities rather than exclusive. In this scenario, multiphoton emissions also contribute to $G^{(2)}(t_1, t_2)$ giving a giant blob of second-order coherence (Fig.~\ref{figure:A1}). The blob has zero values for the equal time correlations $G^{(2)}(t_1, t_1)$ since the two-level system only can emit one photon at a time. Notably, when the pulse is long it inherits the coherence of the incident laser beam and $g^{(2)}[0]=1$. Please see Ref. \cite{Fischer2016-pl} for a discussion on the relationship between temporally-resolved and pulse-wise second-order coherences.


\section*{APPENDIX C: PULSE-LENGTH PHOTOCOUNT DISTRIBUTION FOR 2-PI AREA PULSES}

\begin{figure}[]
  \includegraphics[width=\columnwidth]{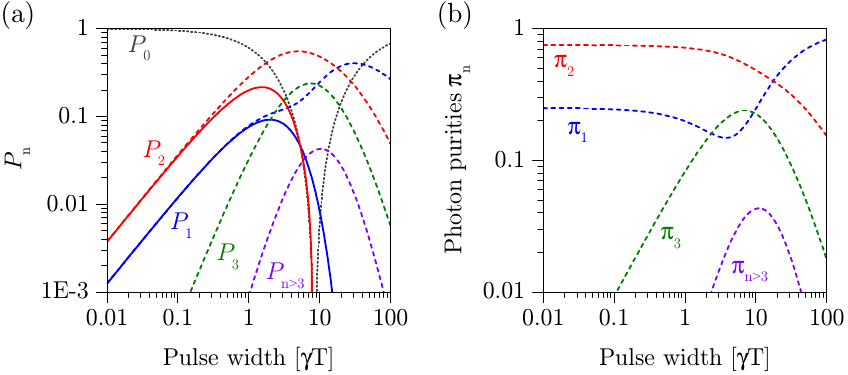}
  \caption{Photon distributions for two-level system under interaction with a square $2\pi$-pulse, i.e. $A(\infty)=2\pi$. (a) Photocount distribution, calculated numerically (dashed curves) and analytically $P_1$ (solid blue) or $P_2$ (solid red). (b) Photon purities $\pi_n=P_n/\sum_{m>0} P_m$, showing highly pure two-photon emission over a large range of pulse widths.}
  \label{figure:A3}
\end{figure}

In this appendix, we investigate the pulse-length dependence of the photocount distribution in the region where two-photon emission dominates. For the square pulse when $A(\infty)=2\pi$, we can arrive at very simple expressions for the probabilities to emit one or two photons. Again, we take $P_1=F_1-F_2$ and $P_2=F_2-F_3$, and get
\begin{eqnarray}
P_1(A(\infty)=2\pi)&\approx&\text{e}^{-\gamma T/2} \frac{\gamma T}{8}\\
P_2(A(\infty)=2\pi)&\approx&\text{e}^{-\gamma T/2} \left(\frac{3\gamma T}{8 } - \frac{3\left(\gamma T\right)^2}{64}\right) \nonumber.
\end{eqnarray}
These analytic functions are plotted in Fig. \ref{figure:A3}a as the solid curves. We also numerically computed the $P_n$ up to three or more photon emissions exactly (dashed curves). The analytic curves match well up to around $\gamma T\approx1$, with $P_1$ and $P_2$ being underestimated. Ideally, we would experimentally work in the region of highest two-photon emission rate $P_2$ that does not spoil its purity with other numbers of photon emissions. One metric for purity is given by 
\begin{eqnarray}
\pi_n\equiv \frac{P_n}{\sum_{m>0}P_m},
\end{eqnarray}
which is a renormalization of the probabilities without the vacuum component \cite{fischer2017signatures}. The purities are only computed numerically (Fig. \ref{figure:A3}b) and show that the two-photon purity remains high to surprisingly large values of the pulse width. Specifically, the two-photon purity begins to drop only as $\gamma T$ approaches $1$. Hence, we could operate in this region and achieve $\pi_2\approx0.71$ and $P_2\approx 0.25$ for an ideal two-level system. Though, we expect even better purities can be achieved with optimization of the pulse shape.

\section*{APPENDIX D: THREE-PHOTON EMISSION PROBABILITY DENSITY}

\begin{figure}[]
  \includegraphics[width=\columnwidth]{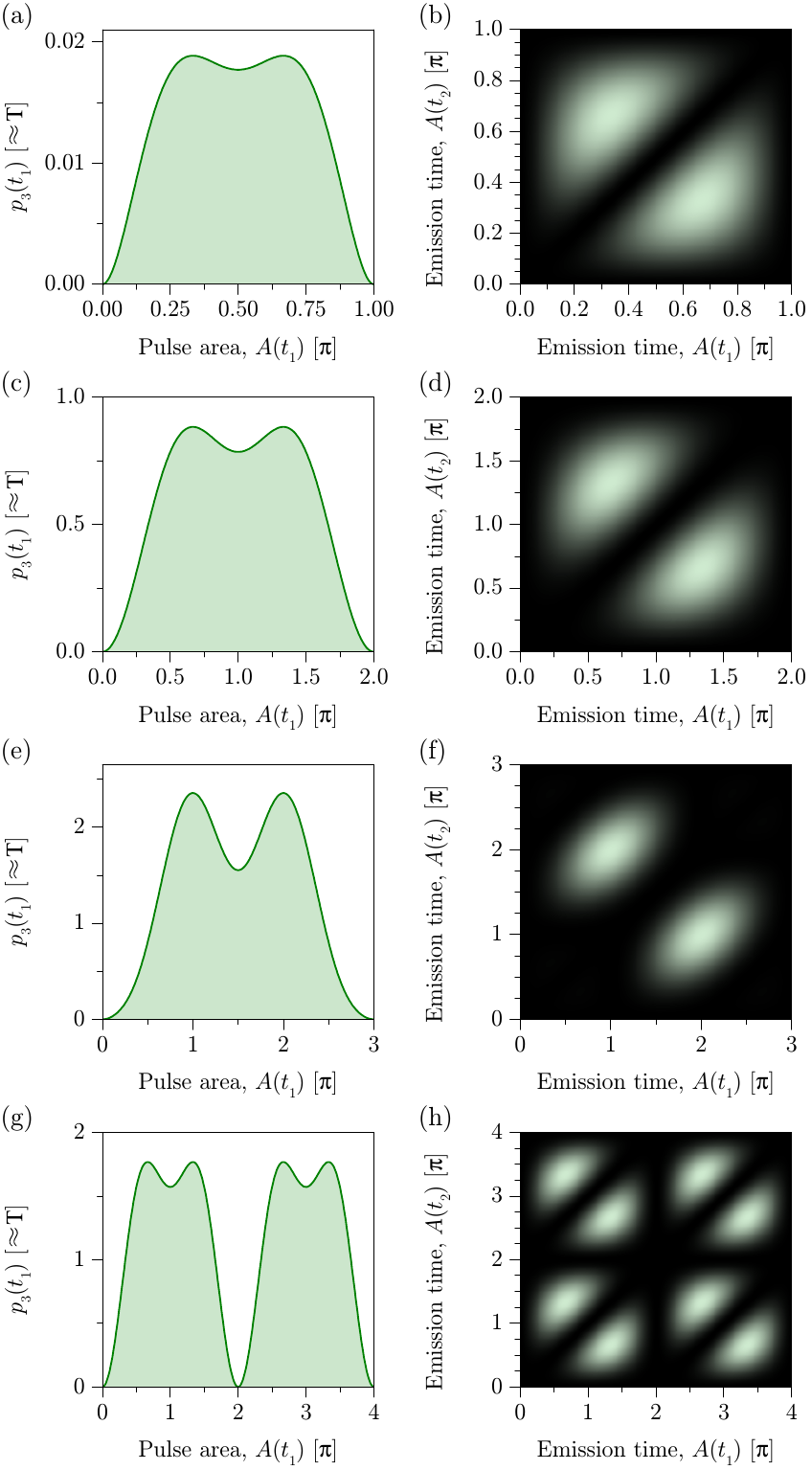}
  \caption{Three-photon emission dynamics for two-level system under interaction with $m\pi$-pulses, i.e. $A(\infty)=m\pi$ for $m\in\{1,2,3,4\}$. (a,c,e,g) Three-photon emission probability density $p_{3}(t_1)$ at time $t_1$, i.e. all the ways three-photon emission might contribute a photon at the time $t_1$, $E[M(t_1)]$. Note, $p_{3}(t_1)$ is not dimensionless. (b,d,f,h) Symmetric three-photon emission probability density $p_{3,\text{S}}(t_1, t_2)$---along the third time axis $t_3$, the probability density $p_{3,\text{S}}$ decays exponentially like $p_2$ does along~$t_2$.}
  \label{figure:A2}
\end{figure}

In this appendix, we explore the possibility for three photoemissions to occur during drive by a short pulse. This scenario with three emissions requires two emissions to occur during the pulse width $ T $, and hence the total probability for three emissions $P_3$ is of the order $( \gamma T)^{2}$. Taking Eq. \ref{eq:fngeneral} for the case of three photon emissions at times $t_1<t_2<t_3$, the inclusive probability density is given by
\begin{align}
f_3(t_1,t_2,t_3)\approx\gamma^3\,\textrm{sin}^2(\tfrac{A(\infty)-A(t_2)}{2})\,\textrm{sin}^2(\tfrac{A(t_2)-A(t_1)}{2})\cdot\nonumber\\
\textrm{sin}^2(\tfrac{A(t_1)}{2})\,\text{e}^{-\gamma t_3}.
\end{align}
Here, we have additionally taken the limit as $T\rightarrow 0$ so we neglect the exponential factor $\text{e}^{-\gamma T/2}$ in our general solution. Further in this limit, the four-photon density is smaller by $\gamma T$ and hence the inclusive and exclusive densities are roughly equivalent
\begin{align}
p_3(t_1,t_2,t_3)\approx f_3(t_1,t_2,t_3).
\end{align}
We will also make use of the symmetric probability densities, which are defined as
\begin{align}
p_{3,\text{S}}(t_1,t_2,t_3)=\frac{p_3(t_1,t_2,t_3)}{3!}+\frac{p_3(t_2,t_1,t_3)}{3!}+\cdots
\end{align}
by adding together $p_3$ with all of its indices permuted. This gives a nice form for the marginal probability densities
\begin{equation}
p_{3,\text{S}}(t_1,t_2) = \int_{0}^\infty \mathop{\textrm{d} t_3} p_{3,\text{S}}(t_1,t_2,t_3),
\end{equation}
and
\begin{equation}
p_{3,\text{S}}(t_1) = \int_{0}^\infty \mathop{\textrm{d} t_2} p_{3,\text{S}}(t_1,t_2).
\end{equation}
These densities are defined for any $t_1>0$ or $t_2>0$.

Using these densities, we explore a number of points. Along the third detection time axis $t_3$, the probability density decays exponentially and is rather uninteresting. The remaining densities have interesting behavior; consider $p_{3,\text{S}}(t_1,t_2)$ as plotted in Fig.~\ref{figure:A2}b,d,f,h for $A(\infty)=m\pi$ and $m\in\{1,2,3,4\}$. These densities are almost all identical in shape for $m\in\{1,2,3\}$. The similarity of the shape owes to the fact that until the interaction area is greater than $3\pi$, the optimal way to achieve a three-photon emission is to divide the pulse into thirds---the increase in excitation probability is then monotonically increasing between emissions. This division is reflected in the two bright regions of the densities, which correlate emissions around $A(\infty)/3$ to those around $2A(\infty)/3$. Meanwhile for $4\pi$, enough area is present that the optimum division must include Rabi flopping between the ground and excited states, resulting in an increase in the number of optimal ways to achieve three-photon emission.

Next, consider the densities $p_{3}(t_1)$ as plotted in Fig.~\ref{figure:A2}a,c,e,g for $A(\infty)=m\pi$ and $m\in\{1,2,3,4\}$. These densities can easily be understood as all possible ways to start a three-photon emission sequence at time $t_1$. Their integrals yield $P_3$ and are depicted as the shaded regions (though the exact values are determined by the specific pulse shape). Notably, $P_3$ is almost two orders of magnitude lower for $A(\infty)=\pi$, which occurs because not enough interaction area is yet present to significantly excite the system between any of the photon emissions, i.e. $\textrm{sin}^2((2\pi/3)/2)^3 \gg \textrm{sin}^2((\pi/3)/2)^3$. As a final remark, we note the connection between photon flux and probability densities. Generally, as long as the marginal densities for photoemission $p_n(t)$ are constructed as exclusive events, then
\begin{equation}
\gamma\,\textrm{E}[M(t)]=\sum_n p_n(t)= \sum_n p_{n,\text{S}}(t),
\end{equation}
which represents all possible ways to achieve a photon emission at time $t_1$. Specifically, $\textrm{E}[M(t)]$ is the expected number of trajectories that contribute an emission at time $t$.


\bibliographystyle{apsrev4-1-PRX}
\bibliography{bibliography}
\end{document}